# Quantum entanglement of angular momentum states with quantum numbers up to 10010

R. Fickler[1,2,3,*], G. T. Campbell[4], B. C. Buchler[4], P. K. Lam[4], A. Zeilinger[1,2,*]

[1] Institute for Quantum Optics and Quantum Information (IQOQI), Austrian Academy of Sciences, Boltzmanngasse 3, 1090 Vienna, Austria
[2] Faculty of Physics, University of Vienna, Boltzmanngasse 5, 1090 Vienna, Austria
[3] Department of Physics and Max Planck Centre for Extreme and Quantum Photonics, University of Ottawa, Ottawa, K1N 6N5, Canada
[4] Centre for Quantum Computation and Communication Technology, Research School of Physics & Engineering, The Australian National University, Canberra, ACT 2601, Australia

*Correspondence to: rfickler@uottawa.ca or anton.zeilinger@univie.ac.at

**Photons with a twisted phase front carry a quantized amount of orbital angular momentum (OAM) and have become important in various fields of optics, such as quantum and classical information science or optical tweezers. Because no upper limit on the OAM content per photon is known, they are also interesting systems to experimentally challenge quantum mechanical prediction for high quantum numbers. Here, we take advantage of a recently developed technique to imprint unprecedented high values of OAM, namely spiral phase mirrors (SPM), to generate photons with more than 10,000 quanta of OAM. Moreover, we demonstrate quantum entanglement between these large OAM quanta of one photon and the polarization of its partner photon. To our knowledge, this corresponds to entanglement with the largest quantum number that has been demonstrated in an experiment. The results may also open novel ways to couple single photons to massive objects, enhance angular resolution and highlight OAM as a promising way to increase the information capacity of a single photon.**

Photonic systems are an excellent platform to test the foundations of quantum physics [1]. Photonic technologies used to generate, manipulate, and measure one of its key features, quantum entanglement, have matured to an unprecedented level. In various experiments the limit of extending quantum mechanical predictions to the macroscopic regimes have been investigated by increasing the distance between the entangled photons [2,3], the numbers of involved photonic systems [4], or the dimensionality of the entanglement [5]. In another approach, the property of orbital angular momentum (OAM), which is related to the helical phase structure of the photons [6] in the paraxial regime and can be used to rotate particles around the optical axis [7], has been explored in quantum entanglement experiments [8]. Interestingly, according to quantum theory this angular momentum can, in principle, be arbitrarily large even for entangled quantum states of single photons. If large enough, this angular momentum can conceivably be transferred to macroscopic particles, which could open novel ways of investigating light-matter interactions. Additionally, large OAM quanta of photonic quantum states are an interesting property to investigate the fundamental question regarding the existence of a quantum-classical transition. It is still believed by many people and often cited in textbooks in connection to Bohr's correspondence principle (see e.g. [9]), that the quantum-classical transition has to occur when the quantum number of the investigated state becomes very large. However, in the opinion of the authors and many others, such a simple relation is not correct (e.g. see also [10]). Therefore, generating quantum states with large quantum numbers, e.g. high OAM values, are important experimental tests to challenge and clarify these fundamental questions. A recent experiment towards these directions demonstrated the entanglement of two photons with up to 300 quanta of OAM [11].

Here, we extend the experimental investigation of entanglement of high angular momenta and report the generation of entanglement between polarization and more than $10^4$ quanta of OAM. To generate this complex hybrid-entangled state, we start with bipartite polarization entanglement and transfer one photon's polarization to a transverse spatial mode carrying 500, 1000 or 10010 quanta of OAM. We realize this transfer in a Michelson-type interferometric scheme, in which we insert spiral phase mirrors to imprint the high OAM quanta onto the photons. We test the generated entanglement in two different ways. First, we use an intensified CCD camera in a coincidence-imaging scheme to show entanglement between the polarization of one photon and its entangled partner that carries 500ħ of OAM. For entangled photons carrying up to 10010ħ of OAM, we replace the camera by an appropriate mask, and again detect OAM superposition states in coincidence with the polarized partner photon. All of the results show clear signatures of quantum entanglement, which demonstrates that OAM quantum numbers of single photons can exceed 10,000 and still be entangled to its partner photon's polarization.

Single photons in a transverse spatial mode with a helical phase front $e^{il\theta}$ ($l$ being an integer value) contain, in the paraxial limit, $l$ quanta of OAM with respect to optical axis [6]. Such phase structures are also often described by its topological charge, namely how many twists the phase of light undergoes in one wavelength. Light exhibiting this twisted phase distribution have a phase singularity, i.e. optical vortex, for $l \neq 0$ in the centre of the mode which leads to a null intensity along the beam axis. Consequently, they are often called "doughnut modes" and have become important for various applications in quantum and also in classical experiments [12]. There are different techniques to generate such phase structures among which computer generated holograms [13], spiral phase plates [14], spatial light modulators [15] or specialized liquid crystal devices, so-called q-plates [16], have attracted much attention.

Recently, a technique was established that is able to imprint unprecedented high quanta of OAM onto light [17,18]. Here, the required spiral phase structure is mapped onto the surface profile of a mirror, hence the name spiral phase mirrors (SPM). To realize the required phase modulation, the surface structure needs to have an azimuthal depth dependence $d(\phi) = l\phi\lambda/(4\pi)$, where $l \in \mathbb{Z}^+$, $\phi \in [0,2\pi)$ is the azimuthal angle and λ corresponds to the wavelength of the photons. If photons are reflected from the surface they experience an azimuthal path difference of $2d(\phi)$ which in turn leads to a spiral phase of $\theta = 2d(\phi)/\lambda$. Thus the imprinted azimuthally-varying phase front corresponds to $l$ multiples of 2π and thus $l$ quanta of OAM. The SPMs are produced by direct machining of the surface of an aluminium disk (2 inch diameter) with an ultra-precision single-point diamond turning lathe (Nanotech 250UPL). For large values of $l$ a steep discontinuity would appear at $\phi = 0°$, which would be an impractical surface to machine. Instead, the surface is divided into $n$ angular segments, each imprinting the n[th] part of the required spiral phase ramp. We are using three different SPMs in our experiments that are fabricated with 25, 25 and 125 segments, each imprinting an azimuthal phase ramp of 40π, 80π or 160π. Hence, photons that are reflected from these SPMs have a spiral phase front corresponding to 500ħ, 1000ħ or 10,000ħ of OAM, respectively (see fig.2 lower right side). Unlike pixelated programmable spatial light modulators, the efficiency of mode conversion from a fundamental Gauss mode to an OAM carrying vortex mode using SPMs is close to that given by the reflectivity of aluminium, provided the machining is perfect and the numerical aperture of the optics is sufficiently large to capture all of the reflected light. With this technique it is possible to shape the wave front of the light with a higher accuracy and generate photons with larger quantum numbers than with any commercially available devices.

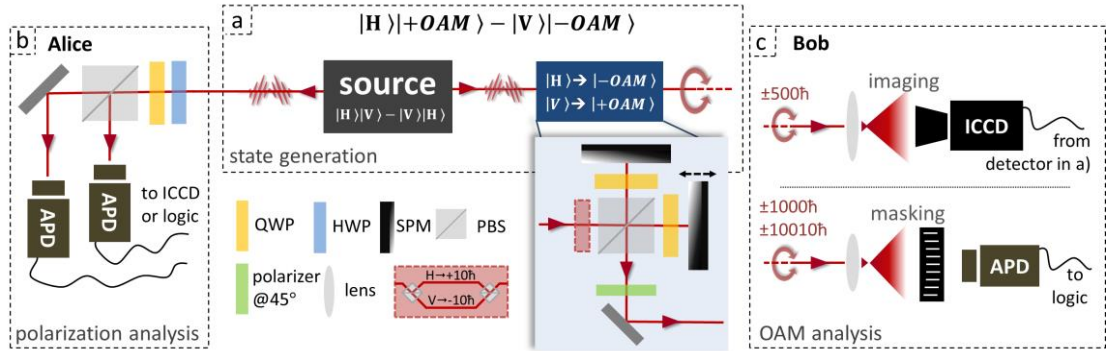

**Figure1:** Sketch of the setup to generate hybrid-entanglement between polarization and high OAM quanta. a) First, polarization entanglement is created (grey box), then one photon is transferred to high OAM quanta in a Michelson-type interferometric scheme (blue box and inset) depending on the polarization. The transfer to ±500 and ±1000 quanta of OAM is done directly from Gaussian modes by spiral phase mirrors (SPM). For the highest quantum number we first transferred photons to ±10 (red dashed box) [11] before sending them into the transfer setup with SPMs for 10,000ħ, thus we generate modes with ±10,010 quanta of OAM per photon. The transfer is completed by projecting on diagonal polarization with a polarizer after the setup. b) Alice: The polarization state of unchanged photon is measured with a combination of half and quarter wave plates (HWP, QWP), a polarizing beam splitter (PBS) and avalanche photo detectors (APD). c) Bob: If the photon is transferred to ±500ħ, we demonstrate entanglement by coincidence imaging, where an intensified CCD camera (ICCD) is triggered depending on the polarization measurement of Alice [25]. If the photons are transferred to ±1000ħ and 10,010ħ, we test entanglement by measuring correlations (logic) between the polarization measurements of Alice and OAM superposition measurements, realized by appropriate masking in front of a bucket detector (APD), of Bob.

The ability to efficiently produce very high-order vortex modes enabled us to test if such high angular momentum quantum numbers can experimentally be imprinted onto light. More importantly, we are able to investigate if single photons carrying high OAM quanta can be entangled with a partner photon's degree of freedom thereby generating quantum states with unprecedented large quantum numbers. The general idea, similar to experiments reported earlier [11, 19-22], is to start with polarization entanglement, send one photon in its unchanged polarization state to Alice and transfer Bob's photon to the OAM degree of freedom while maintaining the entanglement (see fig.1). The transfer of Bob's photon to a mode with high OAM quanta is realized in a Michelson-type interferometer where the photon's path is split dependent on its polarization. Then, the photon's transverse spatial phase profile is modulated by the SPMs to have the high-order helical structure exhibiting a large OAM quantum number. Afterwards, the paths are recombined and the polarization information is erased by a polarizer at 45°. Hence, the created hybrid-entangled bi-photon state shared between Alice and Bob reads as follows

$$|\psi\rangle = a|H\rangle|+l\rangle + e^{i\varphi} b|V\rangle|-l\rangle \quad , \qquad (1)$$

where a, b, and φ are real ($a^2+b^2=1$), $l$ stands for the quanta of OAM, H (V) corresponds to horizontal (vertical) polarization and the positions of the ket-vectors label Alice's and Bob's photon, respectively.

To verify the generated entanglement, we take advantage of a specific feature of the superposition structure of the equally weighted OAM superposition states $|+l\rangle + e^{i\vartheta}|-l\rangle$. The intensity pattern shows a paddle-like pattern of $2l$ maxima in a ring (see fig.2). The orientation $\gamma$ of the pattern is directly linked to the phase $\vartheta$ via $\gamma = \frac{\vartheta}{2l} * \frac{360°}{2\pi}$. Thus, a precise measurement of the paddle structure or the angular probability distribution of Bob's photons, can be directly used to distinguish different superpositions. In combination with the coincident measurement of the polarization of Alice's photon, these measurements can be used to test for entanglement.

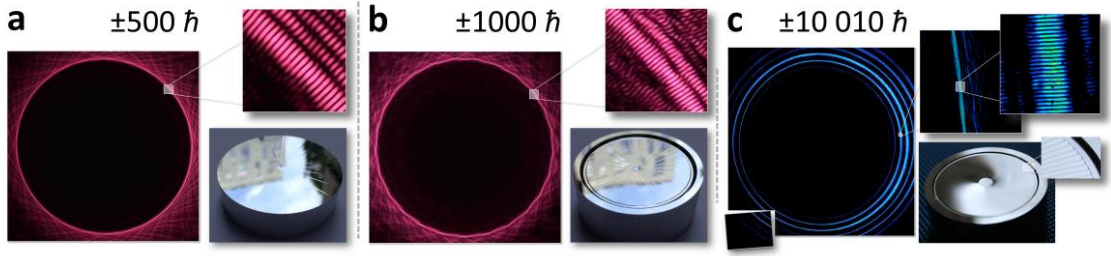

**Figure2:** High-order OAM mode generation with spiral phase mirrors (SPM). Modes are generated by a diagonally polarized laser sending through the transfer setups with SPMs for $500\hbar$ (a), $1000\hbar$ (b) and $10,000\hbar$ (c). Images were taken with a SLR camera ( (a) and (b) ) or by stitching 20 images from a CCD camera together (c). A zoom reveals the typical superposition structure from which the OAM content can be deduced (see main text). Because we only modulate the phase of the light, higher order radial modes with the same OAM content can be found in all images. On lower right sides, photos from the utilized SPMs are shown which are made out of 2 inch aluminium plates. The required helical phase structure is realized by 25 (a), 25 (b) or 125 (c) segments (visible in all photos especially in the zoomed inset in (c) ) each modulating up to $40\pi$, $80\pi$ or $160\pi$, respectively.

In a first series of experiments we test the Michelson-type interferometric transfer setup with a diagonally polarized laser (810nm wavelength) and show that the SPMs are indeed able to generate light fields with unprecedented quanta of OAM. Note, that we also refer to quanta of OAM in the following experiment that only involve light from a laser although it might be perfectly described in the classical Maxwell-theory of light and large topological charges. However, we demonstrate in the experiments investigating quantum entanglement afterwards, that the description with quantum numbers or quanta of OAM is fully justified.

We start by transferring photons from a Gaussian mode with a beam diameter of around 1 inch to a mode with 500 or 1000 quanta of OAM. Typical conversion efficiencies, i.e. percentages of light transferred from the Gaussian mode to OAM carrying light modes, significantly larger than 50% are observed. This becomes apparent in figure 2 due to no visible intensity along the optical axis. In both cases the complex superposition structure formed by $2l$ maxima, which were counted directly from the recordings[1], agrees exactly with the expected values (see fig. 2(a) and (b)).

However, an incident $0^{th}$-order Gaussian mode is unable to realize the transfer to modes carrying $10,000\hbar$. This is due to a difficulty in machining the complex phase profile in the central region of the SPMs. The surface structure corresponding to quantum number of 10,000 is too intricately complex and is therefore not cut (see fig. 2(c)). In order to obtain sufficient mode fidelity and efficiency of conversion to this very high-order OAM modes, we first transfer the photons to $l = \pm 10$ quanta of OAM

---

[1] The maxima were counted by a computer program after contrast enhancement. The proper functioning of the code was tested by re-counting the maxima by hand for the $\pm 500\hbar$ structure.

(depicted in fig. 1 in the figure legend). Because the radial increase of the amplitude for higher order OAM modes scales approximately with $\sim r^{|l|}$, we prevent the light from illuminating the uncut central region and increase the transformation efficiency and mode fidelity. This first transfer is realized with folded Sagnac-like interferometric setup and a spatial light modulator as described in detail in [11], however the polarization information is not yet erased with a polarizer. We then transfer the photons in a second step with the above mentioned Michelson-type setup including the SPMs. We complete the transfer with a polarizer at 45°, such that the photons are in a superposition between +10,010 and -10,010 quanta of OAM. Due to limited resolution of the camera the whole ring-shaped mode had to be imaged by stitching together 20 single images (see fig.2(c)). Moreover, the superposition structure became visible only after further propagation and recording of an even smaller part of the mode, thus simple counting of all maxima to verify the modal order was no longer feasible. Instead, we measured the change of intensity at a fixed angular position while rotating one SPM, thus modulating the phase between $+l$ and $-l$. To account for misalignments and vibrations of the high precision rotation stage, we recorded 5 intensity fringes at 12 different angular positions of the SPM (see supplementary for more details). The average over the twelve slightly different OAM quanta correspond to an equally weighted superposition with $\pm l$ = 9737 with a 1-sigma standard deviation of 327. Thus, we are demonstrating that the SPMs are indeed able to generate light modes with the expected, very high order.

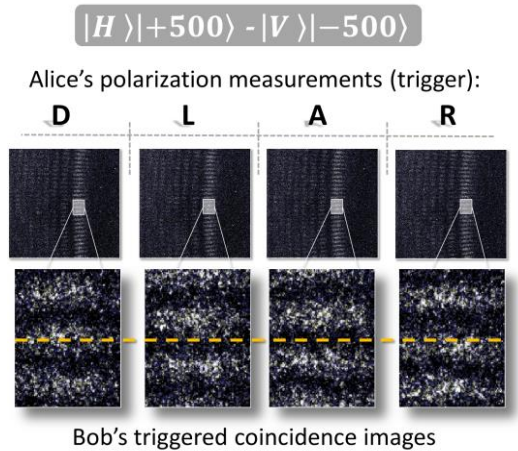

**Figure3:** Coincidence images recorded with a triggered ICCD camera to demonstrate hybrid entanglement of polarization and 500ħ of OAM. Black letters depict Alice's trigger polarization with which the images were recorded (Diagonal, Anti-diagonal, Right- and Left-handed circular polarization). The images, recorded at Bob's location, are built by summing over 60 images each exposed for 10 seconds with a triggering rate of around 500 kHz. A zoom shows the expected shift of the fringes (dashed line to guide the eye) corresponding to different OAM superpositions and depending on the trigger polarization. The opposing location of the extrema of A and D or R and L, respectively, are a direct observable signature of quantum entanglement (see main text).

After having thus demonstrated the transfer classically, we show that an entangled quantum state with these high OAM quanta can be generated and measured. At first, we generate polarization entangled pairs by pumping a type-II down-conversion process (15 mm long periodically poled KTP crystal) with a laser operating at 405 nm (~30 mW power) in a Sagnac-type arrangement [23,24]. For spatial filtering we couple the generated photons into single mode fibres after which we detect approximately 1.2 MHz of polarization entangled photon pairs with a wavelength of 810 nm. Then, we transfer Bob's photon with the schemes described above to ±500ħ, ±1000ħ or ±10,010ħ of OAM while keeping the entangled partner photon of Alice unchanged. To verify the successful generation of this state (1), we used two recently introduced methods, coincidence imaging [25] for ±500 quanta and coincidence detection after masking [11] for ±1000 or ±10,010 quanta.

In the first method we triggered an intensified CCD camera (ICCD, Andor iStar A-DH334T-18U-73, ~20% quantum efficiency, effective pixel size 13 μm × 13 μm, max. 500 kHz triggering rate, 5 ns gating/coincidence window) to image in coincidence the transverse profile of the OAM photon at Bob's location. Here, the polarization measurement of Alice's photon, i.e. the signal of a single photon avalanche photo diode (APD), serves as the trigger. Note that because of the intrinsic delay of the ICCD between the trigger detection and the actual gating of the ICCD, we delay the imaged photon by a 35 m long fibre before it is transferred and recorded. The successful generation of the hybrid-entangled state (1) can be seen in figure 3, where the measurements of different polarization superpositions trigger the camera to image different OAM mode superpositions and thus slightly rotated paddle structures on the ICCD. To more quantitatively verify the non-separable nature of our measurements, we evaluate a witness which corresponds to the sum of the two visibilities in two mutually unbiased bases. Because separable states can only have a perfect visibility of 1 in one of these bases, the witness upper bounds the sum of the two visibilities to 1 for separable states. Values above this classical bound demonstrate quantum entanglement [25, 26]. We measured accidental coincidences by shifting the delay of the trigger signal at the ICCD of around 100ns and subtracted them from the count rates (see supplementary for more details). A witness value of 1.626 ± 0.022 (Poissonian counts statistics assumed) demonstrates a successful generation of quantum entanglement with ±500ℏ of OAM.

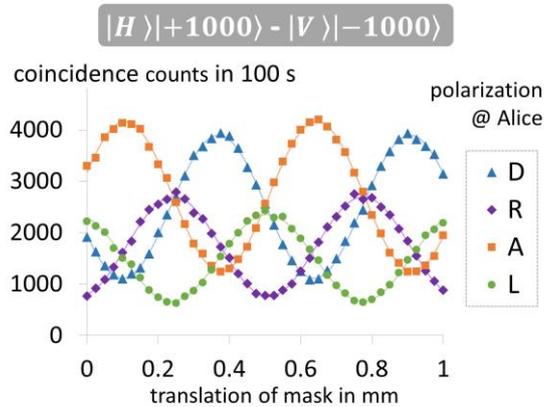

**Figure4**: Coincidence detections between Alice's polarization measurements and Bob's OAM measurements. The OAM photons were measured in different superposition states by detecting transmitted photons behind different positions of a slit mask that mimics the OAM superpositions structure. Because we only used 3% of the full mode and scanned less than two fringes, we approximated the circular rotation of 0.33° by a lateral translation. As expected for an entangled state, extrema of the opposing fringes belong to orthogonal polarizations and cannot be explained by a separable state. The corresponding entanglement witness verifies this observation by more than 10 standard deviations (Poissionian count statistics assumed, error bars are too small to be seen). For measurements in R/L-polarization bases the count rate was smaller, which stemmed from misalignments of the additional quarter wave plate.

Although, the coincidence imaging scheme works very well for SPMs imprinting 500 quanta of OAM, the low triggering rate and detection efficiency of the ICCD prevents its use for higher quantum numbers. Thus, for measurements of entanglement with ±1000ℏ and ±10,010ℏ of OAM we measure the polarization of the unchanged photons of Alice in coincidence with the detections of Bob's transferred photons, which are transmitted through a mask that mimics the superposition pattern [11,27]. The ratio between slit width and the distance between the slits was approximately 1/7 which reduces the maximal measurable visibility to around 97%. Because the masks are fabricated with a low precision laser cutter (resolution approx. 100 µm) and normal paper, we were only able to fabricate and use a small fraction of the whole slit mask (around 60 slits). This allowed us to approximate the curvature of slit arrangement by a linear distribution of slits over approximately 50 mm length. In a similar way we approximated the rotation of the mask during the measurements with a linear translation (see fig4 and fig5). In order to fit the beam of photons to the dimensions of the mask, we had to enlarge the beam size with lenses and free propagation over many meters. The transmitted photons are detected by a bucket detector (APD, active area with a diameter of 500 µm) and recorded in coincidence with 4 polarization measurements of the partner photons (diagonal, anti-diagonal, left and right handed circular). From the shifted extrema of the four measured coincidence fringes, the aforementioned entanglement witness can be evaluated.

For entanglement between polarization and 1000 quanta of OAM we find a value of 1.128 ± 0.013 (Poissonian count statistics assumed) without any subtraction of background or accidental coincidences (see fig.4), which verifies the non-separable nature of the state.

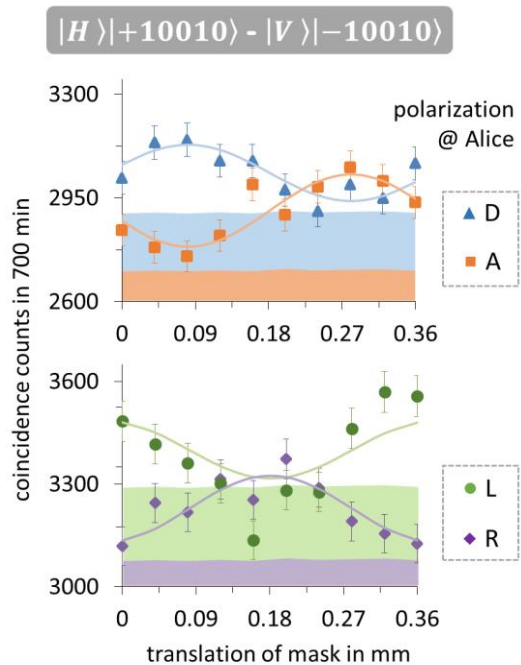

**Figure5:** Quantum entanglement between polarization and 10,010ℏ of OAM. We recorded correlations in the coincidence counts between Alice's polarization measurements and Bob's OAM measurements of superposition states. Again, the OAM superpositions are measured by detecting transmitting photons with a slit mask that mimics the superposition structure. The circular rotation of the mask (<0.018°) is approximated by a linear translation. The coloured region below each fringe corresponds to the accidental coincidence detections, derived from single counts and a measured coincidence window of 4.69±0.34 ns. The error bars show Poissionian error estimation. The lines correspond to the best $\sin^2$-fit function with the condition to be above the accidentals to have non-negative coincidence rates. From the visibilities of the fitted functions, we derived entanglement witness value above the classical bound (see main text).

As is visible in the laser images, the efficiency is drastically reduced for our final step, the generation and verification of entangled quantum states with 10,010 OAM quanta. Various sources of loss, such as the first transfer to ±10 OAM quanta, the bright unused higher-order radial modes (see fig.2(c)), the measurement of only 0.3% of the whole beam and the lower detection efficiency when focusing on the bucket detector after the mask, all lead to a coincidence rate which is more than 10 times smaller than the accidental coincidence detections (see fig.5). Thus, subtraction of accidentals is indispensable. We evaluate them from the measured coincidence window of 4.68 ± 0.34 ns, which is in agreement with the one given by electronics of coincidence logic, and single photon detections (see supplementary). We test for signatures of quantum entanglement by evaluating the witness in two slightly different ways. In the first data analysis, we derive the required visibilities from fitting a $\sin^2$-function thereby taking into account all measured data points and increasing the statistical significance. For fitting we use a least squares method including weighting of data points with respect to the errors and upper bounding the visibility to 1. We evaluate a witness value of 1.43 ± 0.25, which exceeds the classical bound with 1.7 sigma significance. In a second method of analysing the data we split the whole data set into 10 equally long measurement intervals and evaluate for each the witness separately. Again we correct for accidental coincidences by evaluating accidentals from single detections and only take values above the accidentals into account. Here, the mean value and its standard error over all ten measured witnesses is found to be 1.41 ± 0.13 (see supplementary information for more details, e.g. the ten witness values). Both results show a very similar value and are above the classical bound of 1. Thus, our findings can be seen as a clear signature that we successfully generated entanglement between polarization and more than 10,000 quanta of angular momentum.

In summary, we took advantage of the abilities of recently developed spiral phase mirrors to efficiently modulate light into wave fronts with very high complexity. We implemented the SPMs in an interferometric technique that enabled us to generate entanglement between polarization and up to 10,010 orbital angular momentum quanta, thereby extending the measured amount of quanta per photon by two orders of magnitude. As our findings demonstrate a quantum state with a very large quantum number, they can be viewed in the light of discussions about a possible quantum-classical transition for large quantum numbers. Along with the understanding of the authors, our results do not show a signature of such a transition and fully support the quantum mechanical description. Besides these fundamental aspects, the results could also be interesting for questions regarding the maximum capacity of information a single quantum carrier can possess. Although we only generated a two-dimensional state, the results are promising that the spatial degree of freedom of photons offers a very large state space to encode and manipulate quantum information. Similar reasoning also holds true for classical information technology. Moreover, it is known that certain applications like angular sensing benefit in their sensitivity the larger the OAM quantum number is [11,28]. Here, applications where only low light intensities are allowed might especially benefit from the enhancement of large OAM values. Another possible advantage of very high quanta of OAM per single photon might be the enlarged transfer of momenta when light is interacting matter systems, e.g. the coupling of photons to massive objects in quantum optomechanical experiments [29].

**Acknowledgements**

RF thanks M. Krenn for fruitful and motivating discussions and acknowledges financial support by the Canada Excellence Chairs Program and Banting postdoctoral fellowship of the Natural Sciences and Engineering Research Council of Canada (NSERC). The research was funded by the Austrian Academy of Sciences (OEAW), the European Research Council (SIQS Grant No.600645 EU-FP7-ICT), the Austrian Science Fund (FWF) with SFB F40 (FOQUS) and the Australian Research Council Centre of Excellences CE110001027, the Discovery Project DP150101035. PKL is also supported by the ARC Laureate Fellowship FL150100019.

**Author contributions**

RF, AZ and PKL conceived the research; GTC, BCB and PKL designed and fabricated the SPMs; RF designed and performed the experiment; RF and AZ analyzed the data; RF wrote the manuscript with input from all authors.

## Supplementary Information

**Evaluation of 10,010 OAM quanta**

Unlike for the superposition structure of ±500 and ±1000 quanta of orbital angular momentum (OAM), it was not possible to count the expected extrema of the intensity structure for the superposition of $|\chi\rangle = \frac{1}{\sqrt{2}}(|+10010\rangle + e^{i\vartheta}|-10010\rangle)$ directly, because of the limited resolution of the camera. Therefore, we took advantage of the direct relation between a change of the phase $\varphi$ and the rotation $\gamma$ of the intensity structure: $\gamma = \frac{\vartheta}{2l}\frac{360°}{2\pi}$. The idea is to record this rotation while precisely changing the phase $\varphi$ and use the knowledge about the dependence to deduce the OAM content. In our experiment, we realized the phase chance by rotating one spiral phase mirror (SPM) inside the interferometric setup (see fig.1) with a high precision motorized rotation stage (manufacturer-specified absolute accuracy in angular positioning of 0.016 degree). We recorded the rotation of the superposition structure by measuring the intensity at a fixed angular position with a CCD camera while slowly changing the phase, i.e. while precisely rotating the SPM. A change from a measured intensity maximum to a minimum for example would correspond to 1ℏ, 10ℏ, 100ℏ, 1000ℏ and 10,000ℏ if the SPM is rotated by 180°,18°,1.8°,0.18° and 0.018°, respectively. An exemplary measurement for 10,010 quanta of OAM can be seen in figure S1(a). From the $\sin^2$-fit to the measured data points (fig. S1(a) ), we evaluate the OAM quanta within a certain error margin. However, small misalignments of the SPMs or a slight wobble in the rotation causes a squeezing and stretching of the modal structure. Thus, we will record too many or too less fringes than theoretically expected and evaluate an erroneous OAM value. However, when we evaluate the OAM value not only for one small angular region but at different angles we can account such systematic errors. The evaluated 12 OAM values (every 30°) which can be seen in figure S1(b). The squeezing of the fringes, which lead to a larger OAM value than expected, can be seen for the data points up to 150. For the other angular positions of the SPM, a stretching and thus smaller OAM values were evaluated. The average value of 9737 ± 327 agrees within one sigma error with the expected value of a quantum number of 10,010.

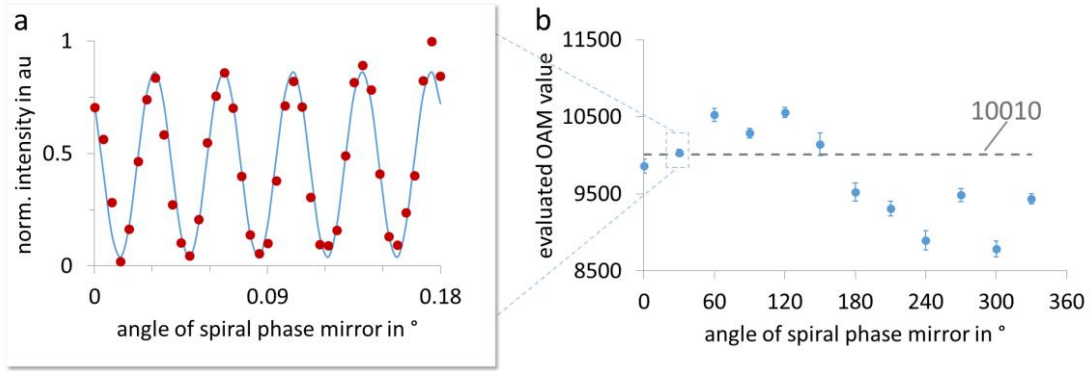

FigureS1: Evaluation of 10,010 OAM quanta. a) The intensity at a fixed transverse position changes depending on a change of the phase $\vartheta$ thus a rotation of one of the two SPMs in the transfer setup (see fig.1 main text). The OAM value can be evaluated from a fit to measured fringe, here 10031 ± 44. b) Different angular position have to be sampled because slight misalignments and wobbles during rotation causing a stretching and squeezing of the modal structure, which would lead to an evaluation of too small or too large OAM values. The average value of 9737 ± 327 agrees with the expected OAM quanta of 10,010.

**Coincidence imaging of entanglement between polarization and 500 OAM quanta**

Coincidence imaging enables the direct visualization of spatial mode entanglement in a nice, intuitive and simple manner. In our experiment for example the shifts of the recorded superposition structures which only depend on the polarization of the partner photon is a direct signature of quantum entanglement. However, for very complex modal structures it has its limitation. Although the resolution is very high, imaging of a superposition structure between +500$\hbar$ and -500$\hbar$ of OAM requires to record only a small part of the beam. Moreover, the limited triggering rate of the ICCD (500 kHz) prevents the use of the full pump power and thus the maximal brightness of the entanglement source. Together with the low quantum efficiency, these restrictions lead to a small overall number of 600 detected photons per image and 10 seconds exposure time. Since the accidentals in such a highly-triggered and long-exposed image were measured be around 200, we had to subtract them from measured coincidence detection. Moreover, we summed up the detected photons for 60 images to get a statistically significant (~30 standard deviations) violation of the witness inequality for separable states. The visibilities we measured were 0.83 ± 0.02 (uncorrected 0.51 ± 0.01) for diagonally/anti-diagonally polarized trigger photons and 0.79 ± 0.02 (uncorrected 0.47 ± 0.01) for circularly polarized triggers. The mentioned limitation hindered a use of the coincidence imaging scheme in the experiment with even higher OAM values.

**Evaluation of entanglement between polarization and 10,010 OAM quanta**

As described in the main text, the very low transformation and detection efficiency for 10,010 OAM quanta requires a subtraction of accidental coincidences. In order to get a good estimation of the accidental rate and its error, we measured the accidental coincidence detections between the transferred and unchanged photons in separate run by delaying the signal of bucket detector after the mask for 100ns. From the detected accidental counts $acc$ and the single photon detections $S_1$ and $S_2$ we can now estimate the coincidence window $\tau_c$ with the help of the formula: $\tau_c = \frac{acc}{S_1 \cdot S_2}$. We found a value of 4.68 ± 0.34 ns, which agrees with the value for which our coincidence logic is specified. With this value and from the measured single photon and pair detections for different mask positions, we analysed our data (shown in fig.5) in two different ways to verify signatures of quantum entanglement.

Method 1: In the first method, we took all measured data points into account and evaluated the visibility of each measured fringe after subtracting for accidental detections. We fitted a sin$^2$-function to our data (Matlab Curve Fitting Toolbox, method 'nonlinear least squares', weights depending on the error) where we kept the expected shift between the fringes and their periodicity fixed. Moreover, we upper bounded the visibilities to be ≤ 1. With this restrictions we obtained values of 1.00±0.47, 0.66±0.15, 0.61±0.13 and 0.58±0.12 for polarized partner photons with right circular (R), left circular (L), diagonal (D) and anti-diagonal (A) polarizations. We then evaluated the visibilities in two mutually unbiased bases (0.83±0.25 for R/L and 0.59±0.09 for D/A) and deduced a witness value of 1.43±0.25. The value exceeds the classical bound of 1 by 1.7 sigma and therefore strongly suggests the verification of quantum entanglement of 10,010 OAM quanta.

Method 2: In a second method of analysing our data, we only took the counts at the extremal points of the four fringes into account. We subdivided our data into 10 equally long measurements (each corresponding to a measurement time of 300 seconds), subtracted accidental coincidence detections and evaluated the witness value for each separately. Again, we limited our values of the visibilities to a meaningful value of 1. This means, that whenever a measured coincidence value was smaller than the calculated accidental coincidence counts, which we deduced from the single photon detections, the coincidence window value and formula given above, we assumed that there were no real pair detections. Thus, in this cases the detected counts were assumed to be zero. After this correction procedure, we obtained ten witness values:

$$0.90 \quad 1.42 \quad 1.98 \quad 0.81 \quad 1.16 \quad 1.26 \quad 1.28 \quad 1.75 \quad 1.68 \quad 1.88$$

The mean value and its standard error, which corresponds to the measure of how far the mean is likely to be from the true mean, is now 1.41 ± 0.13. Again, the obtained value is above than the classical upper bound for separable states of 1, this time with more than 3 standard deviations.

Hence, both methods show clear signatures of the successful generation of quantum entanglement between polarization and 10010 quanta of orbital angular momentum.